# The Advent of Technological Singularity: a Formal Metric.

*Lara, J. A.[1]; Lizcano, D.[1]; Martínez, M. A.[1]; Pazos, J.[1]*

[1]UDIMA-Madrid Open University.

## 1.     Abstract.

The Technological Singularity; that is, the possibility of achieving a General Artificial Intelligence (AGI) that surpasses human intelligence, is one of the vital paradigms of today's humanity. However, until now only opinions about its possibility and/or achievement were issued, therefore, in this work, a metric is presented, for the first time, to objectively measure the actual state in which the advent of technological singularity is found.

## 2.     Introduction.

In the 19th century, machines began to replace humans as physical workers. For many people, this was a blessing, despite the occasional resistance to change led by the Luddites –a secret sect of English textile workers in England who opposed textile machinery-. Another mythical story portraits, in the mid-19th Century, in the USA, a challenge won by John Henry, a Afroamerican with a peak, against a machine, steam-driven, both competing in tunneling. However, after such a pyrrhic victory, Henry died after the effort, clearly anticipating how any tunneling machine greatly exceeds human beings by, several orders of magnitude. Today, no one questions the appropriateness and desirability of replacing humans by machines in these processes.

Today nobody can actually competes with computers to solve "arithmeticly" problems, i.e., those problems that present an effective procedure, algorithmic, to reach a solution. According to the theorem of Löwenhein-Skolen (Löwenheim, 1915) (Skolem, 1920), such a statement is expandable to any normative system, axiomatic or consensual. All formalized *knowledge systems are simply arithmetic.* Thus, for these kind of problems, likewise, it has been proven and accepted that computers outperform humans.

The next level, the authors call *pseudoprogrammed* and, or, heuristic (Pazos, 1987), is that currently begins the controversy and discrepancies between those who are better at solving problems, because it is assumed that require, true intelligence. They are problematic situations which by their vagueness, magnitude, combinatorics, indeterminacy, complexity or other circumstances, are not, *or stricto sens,* algorithmic. Examples of this are the game of poker or, Theorems' demostrations, etc. After all, every time that there are face to face, human *vs* computer, the winner is the computer.

Finally, there is the level of problems and *psicoprogramables* situations, typically involving issues psychological, axiological, teleological, etc., until now considered the highest intellectual level because in their treatment, not only knowledge or intelligence intervene but also , and to a large extent, wisdom. In fact, they are considered, at least up to now, to be specifically human processes. In this sense, the *skeptics technologists* claim that General Artificial Intelligence (AGI) is impossible based on the argument of introspection, on which this type of process is based. However, although the nature of the link between the patterns of electrical signals in neuronal cells and the associated mental processes is still a mystery today, nothing that has been discovered so far suggests and less proof, the presence of something plus exotic, unknown and ignoring that simple electrical and chemical signals that follow well-known physical-chemical laws. On the other hand, as proved by the mental experiment conducted by Rapoport (Rapoport, 1964), computers are much more consistent than humans in the execution of rational decisions.

The main motivation for the development of this research lies in the fact that all the bibliography and research results found about the possibility of pursuing an AGI are *estimates* or opinions (Tegmark,2017), pure "doxagrafía". In this work, a metric is provided to assess what is called the *technological singularity*. Section 2 presents a summary of the *State of Art* regarding technological singularity. In section 3, a proposed solution based on plausible reasoning and Bayes' theorem is presented. Finally, section 4 presents, the results obtained and conclusions drawn.

## 2.    *State of Art*: Technological Singularity.

Contrary to some recent publications (Tegmark, 2017), the term and especially the concept of, "singularity" has a long and substantial history behind. For instance, Condorcet (Condorcet, 1795) was the first to account for a singularity in the human intellectual faculties. Subsequently, Peirce (Peirce, 1867) wondered how much knowledge could be done by machines and how much should be left for humans. Turing, in 1948 (Mitchie, 1998-2004), (Turing, 1948), (Turing, 1950) (Turing, 1952), wrote about intelligent machines. However, the first explicit use, in the technological context, of the term "singularity" was reported in 1957 and its author was von Neumann (Ulam, 1958). It was not until 1988 that Vinge (Vinge, 1988) used the phrase *technological singularity* for the first time, although, he had referred before to the concept (Vinge, 1983). Vinge popularized the term in 1993 (Vinge, 1993) and since then, it has become ubiquitous.

In general and as a concept, *Singularity* means *unique event, Disruptive, with deep and irreversible implications.* For example, in mathematics, involving infinity, in physical black holes, in cosmology *big bang*, and in technology the advent of the *AGI*. This implies that an entity, *cientefacto,* could be improved to itself recursively, in a feedback loop posit that, with probability streaking on certainty, would produce an effect beyond human control. Is obviously will have consequences, anthropological, sociological, economic, etc., unimaginable, impossible to understand and predict for humans.

In order for this technological singularity to occur, three conditions have to be met: A *necessary one*, which is to obtain said AGI, and sufficient ones, namely, that it be *replicated* and made visible. Replication is ensured since von Neumann (Neumann, 1948) revealed to Disraelí, *mystery of mysteries:* the "machines do machines" to provide a process for a computer to make a copy of itself, and that five years before Watson and Crick (Watson, 1953), with the invaluable help of Rosalind Franklin, discovered the secret of reproduction organic, which proved to be practically the same as proposed by von Neumann. And, as far as visibility is concerned, the legend of the discovery of the irrational. (Fritz, 1945), and one of the basic principles of cryptology the authors completed and versified as follows: *one secret, insurance secret | secret of two, keep it God, | secret of three, no secret is, | Four secret, even the cat knows what shows* which, if any, the AGI, would be impossible to maintain the secret.

And there remains the fundamental question of the advent of the AGI. The irrefutable fact is that there are many opinions (Tegmark, 2017) about its viability or not. However, to be known, no one has proposed a procedure that measures whether or not has been achieved, and if not, to what is being achieved. And this is what will be shown below.

## 3.    The proposed solution.

George Polya (Polya, 1954) systematized a series of reasoning patterns that are usually followed in the research processes in the natural sciences, in the creation mathematics, in judicial and police investigations, in the diagnosis medical, etc .:The *plausible reasoning.* Its formal expression is:

If  *A* then *B*

*B*
—————————————————————

*A* is more credible than *B*

Polya, already in 1945 (Polya, 1945), referred to the previous inference as heuristic syllogism, and provided a mathematical model, based on probability, with which to obtain a validation of the rules of knowledge shown in the experience. The first case that addressed Polya, which is what here, now, concerns and interests is the so-called *analysis of a consequence,* and provides two relevant and appropriate rules for what is sought. One, says that the increase of the credit of a hypothesis or conjecture, due to the proof or evidence of one of its consequences, varies inversely to the credibility of the consequence before its test. Two, if *B* without *A* is hard to believe, the verification of consequence *B* leads to the hypothesis or conjecture *A* near certainty. Proof or verification of a result *B* from which there is no doubt even if *A* were false adds virtually nothing to confidence in *A*.

Up to this point, it is clear that if there is evidence about a hypothesis or conjecture that supports it, the hypothesis is more plausible. However, this is merely qualitative, which is why an instrument that quantifies said plausibility is necessary. Fortunately, Bayes (1702-1765) proved a theorem (Bayes, 1763), which today bears his name and offers a method to quantify the plausible reasoning. Later generalized theorem, nostrifying it, Laplace (1749-1827) (Laplace, 1774). In its present form, the rule or Bayes formula, the result of this theorem, is expressed as follows:

$$P(h/e) = \frac{P(e/h).P(h)}{P(e/h).P(h) + P(e/\sim h).P(\sim h)}$$

Where, $h$ represents a hypothesis, conjecture or theory previous or *a priori* abductively inferred before new evidence in the form of facts, observations, etc., resulting available. Its probability is represented by $P(h)$. $P(e/h)$ is the conditional probability that the evidence is met $e$,if the hypothesis $h$ is true. Also called *likelihood function*, when is expressed as a function of $e$ given $h$. $P(e)$ represents the *marginal probability* of,$e$ ,it is, the probability that new evidence be given $e$ under all mutually exclusive hypotheses. $P(h/e)$ it is called *posterior* or *a posteriori probability* of the hypothesis $h$ given the evidence $e$.

On the other hand, the factor  $P(e/h)/P(e)$  represents the impact that the evidence has on the credibility or verisimilitude of the hypothesis. If the evidence is observed when the hypothesis is true, then this factor is big, which multiplied to the previous probability, results in an increase in the posterior probability given the evidence. In short, Bayes' rule measures how much the new evidence can modify the belief in a hypothesis.

The importance of Bayes formula is that it fulfills the formal and material conditions required of any metric adequacy. The first, by its being a theorem, second, because of its simplicity to use a single decision rule and its comparison empirically, again and again, in such disparate and important fields as: paternity tests and identification genetics, cryptography: the *bamburism* procedure used by Turing and colleagues (Good, 1979) in the decipherment of *Enigma*, diagnosis of diseases and malfunctions in mechanical systems, *spam* filters, estimation of submarine itineraries, (McCrayne, 2011) etc.

### 3.1. Sorts, evidences and calculation of probabilities of the model.

Once the evaluation procedure metric is established, it is necessary to provide the existing evidence so that it produces the consequent result. However, it is necessary to be especially demanding and rigorous in the selection of said evidences, because, as is more than known, in any evaluation/information system *If garbage in, garbage out*. In this sense, the authors have taken into account for the appropriate selection of relevant evidence the following:

A) As noted Chalmers (Chalmers, 1999), although confusing *abduction* with *induction*, any hypothesis is corroborated more adequately and better by different kinds of evidence than by one (or some) of a particular class, for the simple reason of the rule of *diminishing returns*. This resulted, for example, that all evidence of games that, according to von Neumann (Bronowski, 1973), are nothing more than calculus like chess, and, more generally, all algoritmizable, as noted in the introduction, are reduce to once "Deep Blue", who beat Kasparov at chess.

B) In all evidence three levels were considered: *Possibility,* which is true, according to von Neuman (Jaynes, 2003) for the simple fact that there is no logical theorem and/or mathematical or law of nature to prevent it, to which is assigned a subjective probability of 50%. *Feasibility*, in the form of a computational system, which raised it probability up to a maximum of 75%. And *Desirability,* for example, when in addition to achieving the goal, the system closed fully the problem facing, as would be the case "Deep Blue" had not only won Kasparov, but would provide an optimal strategy, winning or tables, for chess. Similarly, as Chinook from Shaffler et al. (Shaffler, 1994) was the cheekers' world champion in 1994, and in 2007, its creators proved that the optimal strategy leads to draw.

C) Reciprocally, although sometimes the evidence comes from AI systems of similar name, AlphaGo and *AlphaZero*, by providing different evidence, are considered as such.

D) Classes besides the *features* proposed by Hofstadter (Hofstadter, 1979) were considered. Thus Bayes' rule applied to the evidence(s) in each class, so that in those in which there was no evidence, its value was 50%, as in the case of *proactivity* and integration.

E) Regarding the evidence, the most relevant milestones of the last years, reached by the research in AI have been considered. The authors are aware that a lot of evidence that have not been taken into account for this study are missing, however, the evidences presented here are considered significant and accepted by leading scientists as a sign of facts tested and achieved in IA. It is a flash of the situation in its moment of time and, as a model that it is, in the future it will continue to be nourished and updated with more evidences, such as those related to the camp or music and literary creations., among other.

In Table 1, given both the established sorts for evaluation as evidence considered in this first "experiment".

| | Description |
|---|---|
| **Sorts** | 1. *Holism*: Ability to integrate intelligent elements from a lower level to build a higher level intelligence.<br>2. *Troubleshooting*.<br>3. *Learning*: Acquire knowledge continuously, from all available sources and incorporate them into an integrated and congruent whole.<br>4. *Creativity: Capacity of imagination, intuition and invention.*<br>5. *Teleology: Search for purposes.*<br>6. *Reasoning and inference: Abductive, deductive and inductive, and anterogado.*<br>7. *Proactivity: Initiative to detect interesting problems.*<br>8. *Enantiodromia: Overcoming apparent logical contradictions.*<br>9. *Disambiguation: Overcoming the Turing Test through the challenge of the Winograf schemes.* |
| **Evidences** | 1. *Deep Blue*, chess and *regulated* systems to which the Theorem of Lowenheim-Skolem applies (legislations, regulations, etc.) whose paradigmatic example is the one reviewed *Deep Blue*, that the year 1997 won to Kasparov.<br>2. *Deep Mind* in 2014, found a way to maximize the score in the game of Atari*Breakout*, not predicted by humans, through deep learning, from scratch (Mnih, 2015).<br>3. *AlphaGo*: AI that gave clear signs of intuition and creativity, in its confrontation with the world champion of Go Lee Sedol to which he won, in *movement 37* when placing a chip on line 5, when humans believed the best play in the ranks 3 or 4. Movement 5 confirmed the goodness of the choice of AlphaGo , which also combined, holistically,deep learning and logical GOFAI (D. Silver et al., 2016).<br>4. *AlphaZero*. In 2017, this new AI completely ignored centuries of human experience in Go, including millions of games, learning from scratch, turning fiction into reality by emulating the |

| | | character of Zweig, Doctor B (Zweig, 1943), playing against himself. *AlphaZero*, not only won the *Go* to *AlphaGo*, but also ran in his confrontation with the best chess program. In short, for the first time, the feasibility of AI was improved by improving itself (D. Silver et al., 2017). |
|---|---|---|
| | 5. | *Libratus*. This AI program, Texas Holdem, in 2017, beat four professional poker`s gammer without break, it took 20 days by 14 hours a day, winning $ 1,700,000 of the $ 2,000,000 at stake. The important thing about poker is that the number of moves, superior to chess, is joined by chance in the distribution of cards, and its characteristic of deception and simulation in the bluffs and stakes. |
| | 6. | In 1996, the proof program of theorems EQP (McCune, 1997) proved, creatively, converting it into a theorem, *the Robbins Conjecture*, until then an important open problem. |
| | 7. | In 2011, Watson won the TV *show* ABC *Jeopardy*, facing the two humans who held the record of consecutive victories and accumulated money. The contest lasted three days and the victory of Watson was overwhelming winning three times more money than his opponents. |

Table 1. Sorts and Evidence used for the test experiment.

## 4.     Results and conclusions.

Table 2 shows the total of the probabilities, assigned a priori, to each evidence in each sort, then the Bayes theorem has been applied to them. With these results, a sum of totals was made for each class and for each evidence. Finally, the final probability has been obtained.

| Lessons ╲ Evidences | S1 | S2 | S3 | S4 | S5 | S6 | S7 | S8 | S9 |
|---|---|---|---|---|---|---|---|---|---|
| Ev1 | 0.50000 | 0.85000 | 0.50000 | 0.50000 | 0.75000 | 0.50000 | 0.50000 | 0.50000 | 0.50000 |
| Ev2 | 0.75000 | 0,95775 | 0.90000 | 0.85000 | 0.75000 | 0.50000 | 0.50000 | 0.50000 | 0.50000 |
| Ev3 | 0.92308 | 0.99227 | 0.98077 | 0.96980 | 0.90000 | 0.75000 | 0.50000 | 0.50000 | 0.50000 |
| Ev4 | 0.92308 | 0,99914 | 0,99897 | 0,99836 | 0.98077 | 0.90000 | 0.50000 | 0.50000 | 0.50000 |
| Ev5 | 0.92308 | 0,99985 | 0,99897 | 0.999945 | 0.99351 | 0.90000 | 0.50000 | 0.50000 | 0.50000 |
| Ev6 | 0.92308 | 0.999997 | 0,99897 | 0,99982 | 0.99351 | 0.97297 | 0.50000 | 0.50000 | 0.50000 |
| Ev7 | 0.92308 | 0.999997 | 0,99897 | 0,99982 | 0.99351 | 0,99512 | 0.50000 | 0.50000 | 0.60000 |
| **Final result of the mean probability of the Singularity** | | | | | | | **0.834496158** | | |

Table 2. Probability of the singularity calculated by the Bayes theorem.

In view of these results, the following can be concluded:

1.- With a probability of 83% chance IAG will come true. However, this does not imply that the date on which it will occur can be established by a simple rule of three.

2.- Certainly, So far, virtually all the achievements in AI, were based on digital computers, however, nothing indicates, but quite the contrary, that the IAG as well will be supported exclusively by them.

3.- To make a pertinent analogy, the IAG is currently at the same point as the digital computers before Turing proposed its TMU, there were many particular TMs that solved specific problems, but there was a lack of another TM to integrate them. Said colloquially, all, or almost all, of the wickerwork are available, but the final touch that allows the basket to be made is missing.

## 5. Bibliography.